# Impact of micro-alloying on the plasticity of Pd-based Bulk Metallic Glasses

Niklas Nollmann, Isabelle Binkowski, Vitalij Schmidt, Harald Rösner, Gerhard Wilde

*Institut für Materialphysik, Westfälische Wilhelms-Universität Münster, Wilhelm-Klemm-Str. 10, D-48149 Münster, Germany*

Micro-alloying was performed using additions of Co and Fe to monolithic $Pd_{40}Ni_{40}P_{20}$ bulk metallic glass to study selectively the influence on the plastic behavior in uniaxial compression and three-point bending tests. The corresponding Poisson's ratios were determined by ultrasonic measurements. The microstructure of the individual bulk metallic glasses was characterized by electron microscopy, X-ray diffraction and calorimetry. A plastic strain of 13% was found for the Co addition (1 at.%), whereas the Fe addition (0.6 at.%) led to immediate failure after reaching the elastic limit. Surprisingly, the plasticity is not reflected by the high Poisson's ratio of 0.4 since it remained unaffected by the minor alloying.

Bulk Metallic Glasses (BMGs) show unique mechanical properties such as high strength, extended elasticity, high wear and corrosion resistance[1]. However, the limited ductility and especially the immediate catastrophic failure in tension after reaching the elastic limit are major obstacles for BMGs as structural materials[1]. The limited ductility has led to substantial effort in order to improve the plasticity of BMGs. The design of composites[2–7] composed of ductile crystalline phase in a BMG matrix has been proposed as a promising way to overcome the limited ductility. Moreover, monolithic glasses with high Poisson's ratios near the ideal value of 0.5 are reported to have improved compressive and bending plasticity[8–12]. Thus, the value of the Poisson's ratio has become an indicator for ductile or brittle behavior[9]. In addition, it was found that materials with a high Poisson's ratio form a high number of fine-dispersed shear bands[10,12]. However, small changes in the alloy composition called minor- or micro-alloying were found to cause major changes with respect to the glass forming ability, thermal stability and plasticity of BMGs[13,14]. In this study, micro-alloying was applied to the well-known glass-former $Pd_{40}Ni_{40}P_{20}$ which shows high kinetic stability against crystallization and phase separation[15,16]. Starting from the 'classical' ternary system, micro-alloying was carried out using Co and Fe as additives. Their mechanical response showed completely opposite effects regarding the ductility. The corresponding Poisson's ratios were determined but found to be unaffected by the compositional changes. This raises several questions that need to be addressed: (i) Is the Poisson's ratio the only property that characterizes the ductility? (ii) What hinders the shear bands from propagating through the material leading to catastrophic failure? (iii) What changes are introduced in the glass by minor alloying?

Ingots of $Pd_{40}Ni_{40}P_{20}$ were produced by ingot copper mold casting in a melt spinner under argon atmosphere. The initial alloy compound was modified by adding 1 at.% Co or 0.6 at.% Fe to the ternary system. Before casting, the ingots were cycled with boron oxide ($B_2O_3$) to purify the sample. The sizes of the cylindrical ingots were 3 mm (diameter) x 30 mm (length) for uniaxial compression tests and 30 mm (length) x 10 mm (width) x 1 mm (height) for the three-point bending tests. After casting, the ingots were cut with a diamond wire saw to sample dimensions of 4 mm (length) x 3 mm (diameter) providing a 4:3 aspect ratio for the uniaxial compression tests in accordance with the accepted specification for mechanical tests[17]. The three-point bending experiments were carried out with sample dimensions of 10 mm (length) x 1 mm (width) x 1 mm (height). Uniaxial compression and bending tests were performed in a screw-driven mechanical test instrument (Instron model 1195) equipped with extra hardened anvils made from Böhler S290 microclean steel. The uniaxial compression tests were performed using a strain rate of $2.5 * 10^{-5}$ $s^{-1}$ and the three-point bending tests were carried out at a strain rate of $10^{-3}$ $s^{-1}$. All mechanical tests were repeated several times.

X-ray studies were performed with a Siemens D5000 x-ray diffractometer using Cu $K_\alpha$ radiation to confirm the amorphous state of the samples. In addition, energy-filtered diffraction patterns were acquired with an Omega in-column filter using a slit width of 10 eV to confirm the glassy state locally excluding nano-crystallization. The illuminated area contributing to the diffraction information was 0.5 μm² using an exposure time of 1 s.

Calorimetric measurements were carried out with a differential scanning calorimeter (DSC, TA Instruments DSC Q100) using a heating rate of 20 K/min to monitor the glass transition and crystallization temperature.

The microstructure of the BMG was characterized by optical microscopy (Keyence VHX-500K), scanning electron microscopy (Nova Nano230SEM) and transmission electron microscopy (Zeiss Libra 200FE) using electropolishing with a BK-2 electrolyte[18] for the preparation of electron-transparent samples.

The Poisson's ratios were determined from ultrasonic measurements carried out in an Olympus 38DL Plus device.

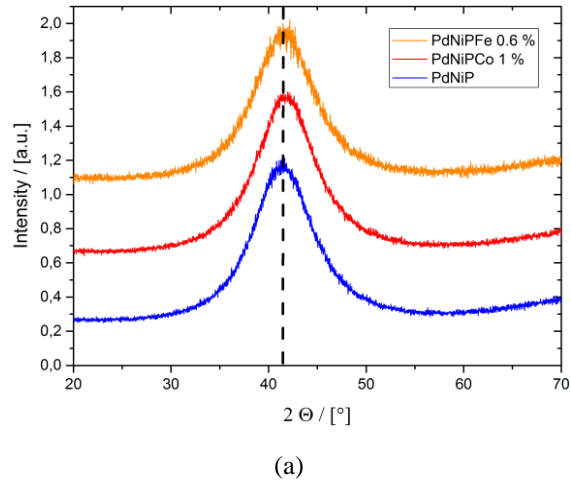

(a)

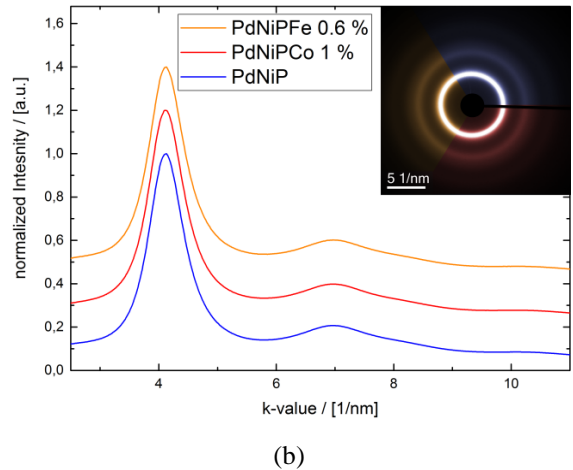

(b)

Fig. 1: (a) XRD pattern of $Pd_{40}Ni_{40}P_{20}$ (red), $(Pd_{40}Ni_{40}P_{20})_{99}Co_1$ (yellow) and $(Pd_{40}Ni_{40}P_{20})_{99.4}Fe_{0.6}$ (blue). (b) Corresponding annular profiles of the rings. The energy-filtered SAED pattern of the different compounds (same color code) is shown as in-set.

Fig. 1(a) shows XRD patterns of different as cast $Pd_{40}Ni_{40}P_{20}$ alloys with and without additions of Co or Fe. They all display the typical X-ray amorphous characteristics of BMGs. Moreover, the glassy structure of the BMGs was also confirmed locally by selected area electron diffraction (SAED) taken from different sample areas (Fig. 1b). No crystalline phases or indications for phase separation were found. Further, as expected for minor additions of Fe or Co, there were no observed shifts in the diffraction maxima of the XRD and SAED pattern.

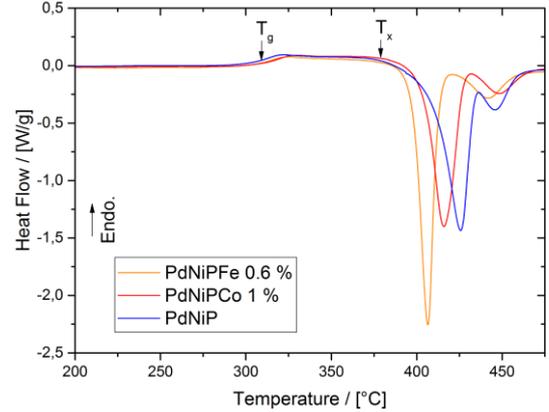

Fig. 2: DSC charts of $Pd_{40}Ni_{40}P_{20}$, $(Pd_{40}Ni_{40}P_{20})_{99}Co_1$ and $(Pd_{40}Ni_{40}P_{20})_{99.4}Fe_{0.6}$ using a heating rate of 20K/min.

| Sample | $T_g$ [°C] | $T_x$ [°C] | $\Delta T$ [°C] | $\Delta H_x$ [J/g] |
|---|---|---|---|---|
| $(Pd_{40}Ni_{40}P_{20})_{99}Co_1$ | 305.9 | 369.2 | 63.2 | 92.8 |
| $(Pd_{40}Ni_{40}P_{20})_{99.4}Fe_{0.6}$ | 305.4 | 370.4 | 64.9 | 91.5 |
| $Pd_{40}Ni_{40}P_{20}$ | 304.2 | 365.8 | 61.6 | 92.4 |

Tab. 1: The enthalpy of crystallization $\Delta H_x$, glass transition temperature $T_g$, crystallization temperature $T_x$, and $\Delta T$ defined as $T_x-T_g$ of the DSC measurements shown in Fig. 2.

The calorimetric results are displayed in Fig. 2 and summarized in Tab. 1. The glass transition temperature $T_g$ (defined as the onset temperature) seems not to be affected by microalloying. However, the crystallization temperature $T_x$ was raised by the minor alloying leading to a slight increase of $\Delta T = T_x-T_g$. Since the onset of crystallization is affected strongly by different factors such as the purity and purification of the material, the cooling rate and melt superheating temperature or, as recently shown[19] by annealing the glass near room temperature, the absolute value of $\Delta T$ bears no significant meaning. The fact that the $\Delta T$- values for the three glass compositions are very similar further indicates that the processing conditions and treatment histories of the materials are equal, which renders their comparison viable.

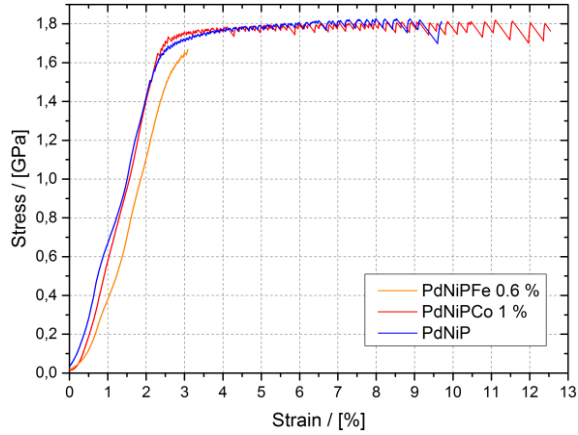

(a)

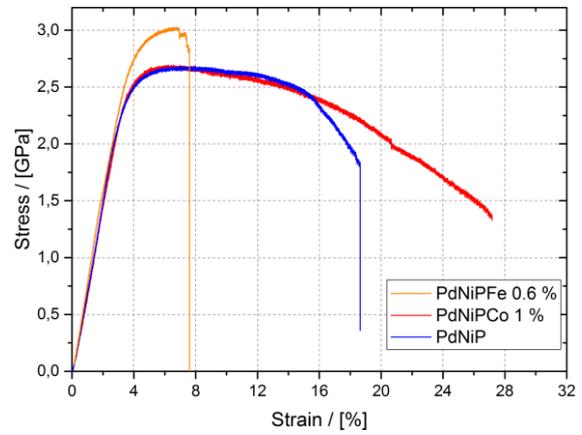

(b)

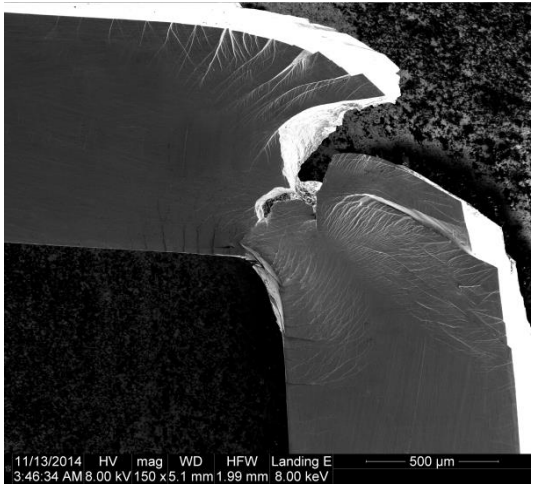

(c)

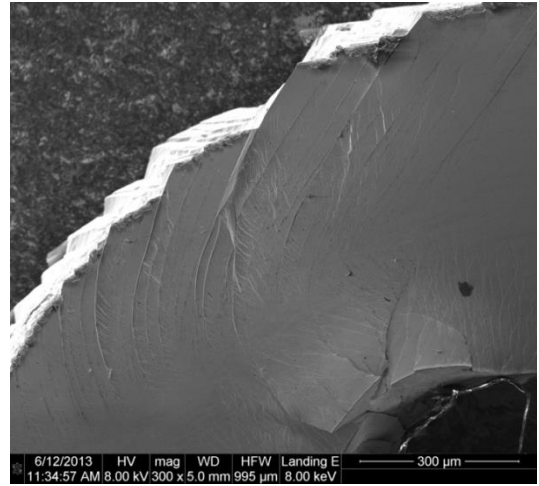

(d)

*Fig. 3: (a) Compressive stress-strain curves of $Pd_{40}Ni_{40}P_{20}$, $(Pd_{40}Ni_{40}P_{20})_{99}Co_1$ and $(Pd_{40}Ni_{40}P_{20})_{99.4}Fe_{0.6}$. (b) Three-point bending tests: stress-strain curves of $Pd_{40}Ni_{40}P_{20}$, $(Pd_{40}Ni_{40}P_{20})_{99}Co_1$ and $(Pd_{40}Ni_{40}P_{20})_{99.4}Fe_{0.6}$. (c) SEM micrograph of a $Pd_{40}Ni_{40}P_{20}$ sample after failure showing shear band details at the bending position. (d) SEM micrograph of bent $(Pd_{40}Ni_{40}P_{20})_{99}Co_1$ sample showing shear band details at the bending position.*

The results of the mechanical testing are shown in Fig. 3. The original ternary $Pd_{40}Ni_{40}P_{20}$ alloy already shows a plastic strain of more than 9% in compression (Fig. 3a). This appears to contradict a former report where only 0.4% of plastic strain was reported for this compound[20]. However, the reported tests were carried out with rectangular shaped samples having an aspect ratio of 3:1 and were thus different in that case. Co addition (1 at.%) leads to an increase in plastic strain of about 50%, whereas the material with Fe addition (0.6 at.%) immediately shows catastrophic failure after reaching the elastic limit. It was shown that the activation of shear bands can be directly linked to the flow serrations in uniaxial compression tests[21]. We thus assume that the formation of each such major shear band is manifested in a single serration during the uniaxial compression test (Fig. 3a).

In addition, three-point bending tests were performed. Videos recording the progress in time-lapse can be found online (see Fig. 4). An increased ductility was found (Fig. 3b) making it impossible to deform the Co containing samples to failure. Fig. 3c shows a SEM micrograph of a $Pd_{40}Ni_{40}P_{20}$ bending sample after failure. There are long primary shear bands visible for the upper part (tensile state) of the bent sample and additionally secondary and tertiary shears bands branching off. However there is one major shear band that, apart from the one that causes the failure, crosses the neutral fiber. The lower part, which displays the compressive state, shows many fine-dispersed shear bands and only very few long shear bands (dark contrast) as found in the upper part. Fig. 3d shows a SEM micrograph of a bent but not broken $(Pd_{40}Ni_{40}P_{20})_{99}Co_1$ sample revealing the differences. There are also long primary shear bands visible for the upper part (tensile state) of the bent sample and additionally secondary and tertiary shears bands branching off. However, none of the long major shear bands crosses the neutral fiber. Thus, there must be a mechanism operating which hinders the major shear bands from propagating through the whole sample. We think that the highly branched state with many secondary and tertiary shear bands involved represents the mechanical response to the material to local modifications of the energy landscape. This highly branched state prevents the primary shear bands from spreading further and thus the material from failure. The lower part, which displays the compressive state, shows again many fine-dispersed shear bands but fewer major shear bands. In contrast, the sample with the Fe addition (not shown) shows no shear bands since it fails with the initiation of the first shear band running through the whole sample.

| Sample | Poisson ratio | Δ Poisson ratio |
|---|---|---|
| $(Pd_{40}Ni_{40}P_{20})_{99}Co_1$ | 0.403 | 0.001 |
| $(Pd_{40}Ni_{40}P_{20})_{99.4}Fe_{0.6}$ | 0.402 | 0.001 |
| $Pd_{40}Ni_{40}P_{20}$ | 0.400 | 0.001 |

*Tab. 2: Poisson's ratios of $Pd_{40}Ni_{40}P_{20}$, $(Pd_{40}Ni_{40}P_{20})_{99}Co_1$ and $(Pd_{40}Ni_{40}P_{20})_{99.4}Fe_{0.6}$ determined from ultrasonic measurements.*

The Poisson's ratios measured are shown in Tab. 2. For the ternary $Pd_{40}Ni_{40}P_{20}$ glass, we obtain a value of 0.400 which correlates well with other measurements[22,23]. However, the Poisson's ratios of the samples containing Co or Fe do not show changes within the measuring error. Thus brittle or ductile behavior can exist at almost identical Poisson's ratios. This result is surprising given the generally accepted finding of Lewandowski et al.[9] that Poisson's ratio is an indicator for plastic and brittle behavior of a bulk metallic glass.

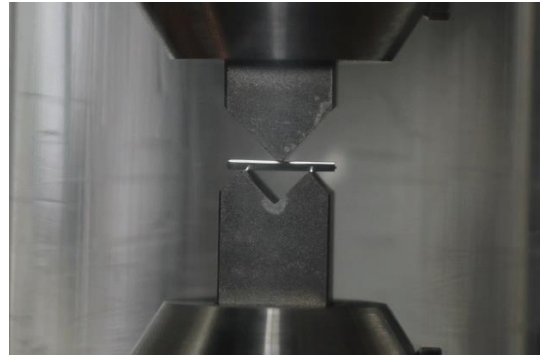

*(http://www.uni-muenster.de/Physik.MP/Biegeversuch.html)*

*Fig. 4: Video of the bending test of $(Pd_{40}Ni_{40}P_{20})_{99}Co_1$.*

Two new $Pd_{40}Ni_{40}P_{20}$ based metallic glasses were developed by micro-alloying, one with an additional content of 1 at.% Co and the other with 0.6 at.% Fe. The $(Pd_{40}Ni_{40}P_{20})_{99}Co_1$ samples showed increased plasticity of about 50% compared to the ternary system in uniaxial compression experiments. On the other hand Fe addition immediately leads to a complete loss of ductility. The ultrasonic measurements did not show any significant change in the Poisson's ratios indicating that the ductility of BMGs is not adequately described by the Poisson`s ratio alone. We observe highly branched states with many secondary and tertiary shear bands involved for the ductile compounds. We think that the complex branching prevents the primary shear bands from spreading further and thus the material from failure.

**Acknowledgment**

We gratefully acknowledge financial support by the DFG via SPP 1594 (Topological engineering of ultra-strong glasses).